\documentclass[pra,amsmath,amssymb,amsfonts,nofootinbib]{revtex4}

\usepackage{bm,graphicx,mathrsfs}
\usepackage{enumerate}
\usepackage{amsmath,bbm}
\usepackage{amsfonts,amssymb}
\usepackage{graphicx}
\usepackage{epsfig}
\usepackage{times}
\usepackage{appendix}
\usepackage{subfigure}
\usepackage{color}

\newcommand{\bR}{\mathbbm{R}}

\newcommand{\1}{\mathbbm{1}}
\newcommand{\id}{{\rm id}}

\newcommand{\half}{\frac{1}{2}}

\newcommand{\cL}{{\cal L}}
\newcommand{\cM}{{\cal M}}
\newcommand{\cS}{{\cal S}}

\newcommand{\tr}{{\rm tr}}

\newcommand{\be}{\begin{equation}}
\newcommand{\ee}{\end{equation}}
\newcommand{\bea}{\begin{eqnarray}}
\newcommand{\eea}{\end{eqnarray}}

\newcommand{\ket}[1]{|#1\rangle}
\newcommand{\bra}[1]{\langle#1|}

\newcommand{\ra}{\rightarrow}

\def\Proof{\noindent\textsc{Proof}}
\def\proof{\Proof}
\def\qed{\leavevmode\unskip\penalty9999 \hbox{}\nobreak\hfill
     \quad\hbox{\leavevmode  \hbox to.77778em{%
               \hfil\vrule   \vbox to.675em%
               {\hrule width.6em\vfil\hrule}\vrule\hfil}}
     \par\vskip3pt}

\begin{document}

\newtheorem{theorem}{Theorem}
\newtheorem{lemma}[theorem]{Lemma}
\newtheorem{corollary}[theorem]{Corollary}
\newtheorem{proposition}[theorem]{Proposition}
\newtheorem{definition}[theorem]{Definition}
\newtheorem{example}[theorem]{Example}
\newtheorem{conjecture}[theorem]{Conjecture}

\newenvironment{remark}{\vspace{1.5ex}\par\noindent{\it Remark}}%
    {\hspace*{\fill}$\Box$\vspace{1.5ex}\par}

\title{\sc{\Large A Cutoff Phenomenon for Quantum Markov Chains}}

\author{Michael J. Kastoryano$^{1,}$\footnote{Electronic address: kastoryano@nbi.dk}}
\author{David Reeb$^{2,}$\footnote{Electronic address: david.reeb@tum.de}}
\author{Michael M. Wolf$^{\,2,}$\footnote{Electronic address: m.wolf@tum.de}}
\affiliation{$^1$ Niels Bohr Institute, University of Copenhagen, 2100 Copenhagen \O, Denmark\\ $^2$ Department of Mathematics, Technische Universit\"at M\" unchen, 85748 Garching, Germany}

\begin{abstract}
We derive upper and lower bounds on the convergence behavior of certain classes of one-parameter quantum dynamical semigroups. The classes we consider consist of tensor product channels and of channels with commuting Liouvillians. We introduce the notion of \emph{Cutoff Phenomenon} in the setting of quantum information theory, and show how it exemplifies the fact that the convergence of (quantum) stochastic processes is  not solely governed by the spectral gap of the transition map. We apply the new methods to show that graph states can be prepared efficiently, albeit not in constant time, by dissipation, and give the exact scaling behavior of the time to stationarity.
\end{abstract}

\date{\today}
\maketitle

\section{Introduction}\label{intro}

The Quantum Information and Computation community has thus far mostly focused on discrete time processes and unitary dynamics. Recently, however, the focus has been shifting towards more physically motivated processes, which inevitably involve continuous time and open systems dynamics. One of the major bottlenecks in understanding the information processing potential of open quantum systems has been the lack of quantitative tools for analyzing their convergence to stationarity. The interest in understanding the convergence behavior of quantum dissipative time-evolutions and quantum channels has cropped up in several distinct areas of quantum information theory, and seems to require a new set of tools for its analysis. The aim of this paper is to present one tool in this direction -- the cutoff phenomenon -- and illustrate it in explicit examples and applications.

As a first step in developing tools for studying convergence rates of quantum processes, it is important to understand how much of the machinery from the classical Markov chain literature can be borrowed. Indeed, by noting that a finite-dimensional quantum channel is the non-commutative analogue of the probability transition matrix of a finite-state Markov chain, many results from the field of Markov chain mixing can be translated to the quantum setting, with appropriate modifications. A first effort in this direction was taken in \cite{Chi2}, where $\chi^2$-convergence was considered, a quantum version of detailed balance was defined, and a restricted quantum Cheegers bound was proved. Furthermore, Hilbert's projective metric \cite{hilbertprojectivem} yields quantities which upper bound convergence rates. These studies and the few other attempts at quantifying the convergence behavior of quantum processes have focused on convergence for asymptotic times.

More generally, it has become almost a folk theorem that ``the gap governs the convergence'' of (quantum) stochastic processes. This statement is true in the sense that the spectral gap represents the exponential convergence rate of a (quantum) stochastic process for sufficiently large times. However, there are situations where the asymptotic behavior either does not kick in before an amount of time exponential in the system size, or does not address the relevant physics. In particular, the important question is often ``after \emph{how much} time, as a function of some dimensional parameter (e.g.~the system size), does the gap properly describe the convergence behavior?'' In the case of state preparation \cite{DSP1}, dissipative quantum computation \cite{DQC1}, or thermalization \cite{Therm1}, it is critical to guarantee that the asymptotic regime is reached in a time which is not exponential in the system size. In the case of quantum memories and error correction, on the other hand, one would like to guarantee the opposite direction, i.e.~that the information initially encoded in the system is preserved for as long as possible, e.g.~for an exponential amount of time.

One of the few situations where one can make rigorous statements about the pre-asymptotic behavior of classical Markov chain convergence is known as the \textit{Cutoff Phenomenon} \cite{cutoff1,cutoff2,cutoff3}. Loosely speaking, the cutoff phenomenon describes the situation where the (quantum) Markov chain, for some initial states, stays far away from its stationary distribution for a possibly long time (thus, e.g.~preserving classical information), and then, at a specific time that may depend on the system size, suddenly approaches the fixed point (thereby suddenly losing all information from the initial state). Cutoff phenomena depend, apart from the type of Markov chains, on the chosen distance measure. The prototypical example of this behavior is in card shuffling \cite{cutoff2}, where a deck of cards is well shuffled only after a number of shuffles logarithmic in the deck size, whereas before that time it has large dependence on the initial ordering. In this article, we state a quantum version of this framework, and apply it to some situations of relevance in Quantum Information Theory.  

The outline of the paper is as follows. In Section \ref{basicnotions}, we set the notation and introduce basic mathematical tools, in particular spectral properties of quantum dissipative time evolutions and distance measures. In Section \ref{cutoffsection}, we give a definition of the Cutoff Phenomenon in the quantum setting. In Section \ref{mainresultssecdt}, we state and prove the main results, namely cutoff-type bounds for time-evolutions due to commuting Liouvillians or tensor product channels. Section \ref{examplessect} illustrates these main results by various examples. As an application, we show in particular that the dissipative preparation of graph states takes a time logarithmic in the system size, and is thus not solely governed by the spectral gap of the Liouvillian. Finally, the conclusion and a brief outlook are given in Section \ref{conclusionsect}.

\section{Basic Notions}\label{basicnotions}

\subsection{Mathematical Setting}
Throughout the paper, we restrict ourselves to finite-dimensional quantum systems. The set of quantum states (density matrices) acting on a $d$-dimensional Hilbert space is denoted by $\cS_d = \left \{ \rho \in \cM_d \big| \rho = \rho^{\dag}, \rho \geq 0, \tr[\rho] = 1 \right \}$, its restriction to full-rank density matrices by $\cS^+_d$. Quantum channels are completely positive trace-preserving linear maps $T: \cM_d\rightarrow\cM_d$, where here the input and output spaces are taken equal to allow for repeated application of the channel. The dual map, i.e.~the channel in the Heisenberg picture, is denoted $T^*$ and by definition satisfies $\tr[AT^*(B)]=\tr[T(A)B]$ for all $A,B\in\cM_d$. We consider almost exclusively quantum channels $T_t:=e^{t\cL}$ (for $t\geq0$) that are elements of the quantum Markov semigroup generated by a Liouvillian $\cL$, which due to the trace-preserving property of $T_t$ satisfies $\cL^*(\mathbbm{1})=0$. $T_t$ describes the evolution after time $t$ coming from the master equation $\dot{\rho}=\cL(\rho)$. When fixing some complete orthonormal basis $\{ F_{i} \}_{i=1, \ldots, d^2}$ of $\cM_d$, any linear map $S:\cM_d\ra\cM_d$ can be represented as a $d^2\times d^2$-matrix $\hat{S}$ by defining $\hat{S}_{ij}:=\tr{[ F_{i}^\dagger  S (  F_{j})]}$. We dress the operator $S$ with a hat when working in the matrix representation.

\subsection{Spectral Properties of Quantum Channels}
The spectrum $\Sigma(T)$ of a quantum channel $T:\cM_d\rightarrow\cM_d$ is contained in the closed unit disk of the complex plane, and every eigenvalue has a complex conjugate partner. One eigenvalue is $1$, the corresponding eigenspace is spanned by quantum states, and the density matrices in this eigenspace are exactly the stationary states of the channel. Like any linear map, a quantum channel can be decomposed uniquely into Jordan normal form:
\be\hat{T}~=~\sum_k \mu_k \hat{P}_k +\hat{N}_k~,\nonumber\ee
where each $\mu_k\in\Sigma(T)$ is an eigenvalue of $\hat{T}$, $\hat{P}_k$ is a projection onto the corresponding Jordan eigenspace, and $\hat{N}_k$ is a nilpotent matrix on this eigenspace. It can be shown that the eigenvalues $\mu_k$ with modulus $|\mu_k|=1$ (the so-called peripheral spectrum) have trivial Jordan blocks ($\hat{N}_k=0$), and thus the asymptotic space of the quantum channel $T$ is $\aleph_T:={\rm span}\{X\in \cM_d \big| \exists \theta \in \bR:T(X)=e^{i \theta}X\}$. Its asymptotic evolution is the phase-preserving projection onto $\aleph_T$,
\be\hat{T}_\varphi~:=~\sum_{k: |\mu_k|=1} \mu_k \hat{P}_k~,\label{periphchannel}\ee
and the corresponding map $T_\varphi:\cM_d\ra\cM_d$ is a quantum channel. If $T_t=e^{t\cL}$ comes from a one-parameter semigroup of quantum channels, the spectra and (generalized) eigendecompositions of $T_t$ and of $\cL$ are related by exponentiation. In this case, we suppress the time-dependence when writing $T_\varphi:=(T_t)_\varphi$.

If a channel $T$ has only one eigenvalue of modulus $1$ and if its unique stationary state has full rank, then we call $T$ \emph{primitive} \cite{Wielandt}. We call a Liouvillian generator $\cL$ primitive iff $e^{t\cL}$ is a primitive channel for all $t>0$ or, equivalently, iff $e^{t\cL}$ is primitive for one $t>0$ or, equivalently, iff $\cL$ has exactly one eigenvalue $0$ and the corresponding stationary state $\rho$ (i.e.~$\cL(\rho)=0$) has full rank.

Note that $(T-T_\varphi)$ has spectral radius equal to ${\bar{\mu}}$ where ${\bar{\mu}}:=\sup\{|\lambda|\,\big|\,\lambda\in \Sigma(T), |\lambda|<1\}$ is the largest modulus of the eigenvalues of $T$ in the interior of the unit disk. If $T_t=e^{t\cL}$ comes from a one-parameter semigroup, then ${\bar{\mu}}(t)=e^{-t \bar{\lambda}}$ where $\bar{\lambda} :=\inf\{|{\rm Re}(\lambda)|\,\big|\,{\rm Re}(\lambda)<0,\lambda\in\Sigma(\cL)\}$, and $\bar{\lambda}$ is referred to as the {\it gap} of the Liouvillian. For a more detailed discussion of spectral properties of quantum channels, see \cite{InverseEIG,Wielandt}.

\subsection{Distance Measures and Contraction Coefficients}\label{distancemeasuresection}
There are many ways to quantify distance between two quantum states. We recall basic relationships between the distance measures used in this paper. For $\rho,\sigma\in\cS_d$ define the distance measures:
\begin{enumerate}
\item The trace distance: $d_\tr(\rho,\sigma)=\frac{1}{2}||\rho-\sigma||_1$, where $||X||_1=\tr[\sqrt{X^\dag X}]$ is the trace-norm.
\item The Bures distance: $d_B(\rho,\sigma)=\sqrt{1-F(\rho,\sigma)}$, where $F(\rho,\sigma)=\tr{[\sqrt{\sqrt{\rho}\sigma\sqrt{\rho}}]}$ is the fidelity.
\end{enumerate}

The trace and Bures distances have the important property that they are monotone under the application of quantum channels, i.e.~$d(T(\rho),T(\sigma))\leq d(\rho,\sigma)$ for any channel $T$ and any states $\rho, \sigma$. It is well known that they satisfy \cite{NielsenChuang}:
\be d_B^2(\rho,\sigma)~\leq~d_\tr(\rho,\sigma)~\leq~d_B(\rho,\sigma) \sqrt{2-d^2_B(\rho,\sigma)}~=~\sqrt{1-F^2(\rho,\sigma)}~.\label{TrNrmFid}\ee
Moreover, for states $\rho$ close to a full-rank state $\sigma$, the Bures and trace distances can be bounded even linearly in terms of each other (see Appendix \ref{appendixa} for a proof):
\begin{proposition}[Trace vs.~Bures distance infinitesimally]\label{burestnLinear}Let $\sigma\in\cS_d^+$ be a strictly positive density operator with smallest eigenvalue $\lambda_{min}(\sigma)>0$. Then there exists $\epsilon=\epsilon(\sigma)>0$ such that
\be d_B(\rho,\sigma)~\leq~\frac{1}{\sqrt{\lambda_{min}(\sigma)}}\,d_\tr(\rho,\sigma)\label{burestrlineareqn}\ee
for all density matrices $\rho\in\cS_d$ with $d_\tr(\rho,\sigma)\leq\epsilon$.
\end{proposition}

Having introduced appropriate distance measures between quantum states, we now want to characterize the distance to stationarity of a quantum channel. Such a convergence measure should somehow quantify one or several of the three closely related properties: {\it{(i)}} how far an output state of the channel $T$ may be from its limiting evolution $T_\varphi$, {\it{(ii)}} how reversible the action of the channel is, {\it{(iii)}} how much information is lost by the application of the channel. We choose the following definition, which addresses point {\it{(i)}} above.
\begin{definition}[Trace-norm contraction]Let $T:\cM_d\rightarrow\cM_d$ be a quantum channel. Then its (trace-norm) contraction is defined as
\be \eta_\tr[T]~:=~\sup_{\rho \in S_d} d_\tr\left(T(\rho),T_\varphi(\rho)\right)\label{ContractionCoeff}~.\ee
\end{definition}
The Bures contraction $\eta_B[T]$ is defined analogously by using the Bures distance $d_B$. Eqn.~(\ref{ContractionCoeff}) is not the only possible choice of a contraction measure, and like the other possibilities it has strengths and weaknesses. We point out in particular that Eqn.~(\ref{ContractionCoeff}) can be discontinuous in $T$, whenever the size of the peripheral spectrum is discontinuous.

In the setting of dissipative state preparation (where the peripheral spectrum is usually engineered to be trivial), $\eta_\tr[T]$ equals the maximum trace distance between the desired steady state and the channel output, i.e.~for the most disadvantageously chosen initial state $\rho$. Having this in mind, we now state an asymptotic convergence theorem for one-parameter semigroups of quantum channels, which is also a crucial fact guiding our further investigations.
\begin{theorem}[Contraction theorem]\label{contractionrate}Let $\cL:\cM_d\rightarrow\cM_d$ be a Liouvillian, i.e.~the generator of a one-parameter semigroup of quantum channels $T_t\equiv e^{t\cL}$ $(t\geq0)$, and let $\bar{\lambda}$ be the gap of $\cL$. Then, there exists $L>0$ and for any  $\nu<\bar{\lambda}$ there exists $R>0$ such that
\be L e^{-t \bar{\lambda}}~\leq~\eta_\tr[T_t]~\leq~R e^{-t \nu}\qquad\forall t\geq0~.\label{contractionrateEqn}\ee
\end{theorem}
A proof is supplied in Appendix \ref{appendixb}, showing in particular that one may not choose $\nu={\bar{\lambda}}$ when an eigenvalue $\lambda_k$ of $\cL$ with modulus $|\lambda_k|={\bar{\lambda}}$ has a non-trivial Jordan block. Theorem \ref{contractionrate} makes a statement only about the asymptotic convergence behavior, i.e.~the exponential rate as $t\to\infty$. If $\nu$ is taken close to $\bar{\lambda}$, then $R$ can become very large. But for fixed $\nu$, a universal dimension-dependent upper bound on $R$ of order $d^{d^2}$ can be obtained. If the system at hand is composed of many particles, say $n$ qubits, then this time-independent prefactor can in principle become doubly exponentially large in the number of particles; this would correspond to an exponentially long time to convergence, even for a gap constant in the system size.

Finally, we note that an analogue of Theorem \ref{contractionrate} holds also for the repeated application of a discrete-time channel $T$, and results analogous to the ones following can be formulated in the discrete-time setting as well.

\section{The Cutoff Phenomenon}\label{cutoffsection}

Often times, the relevant question when considering the convergence behavior of open systems is ``how long does the process have to run before it reaches equilibrium?'' To make precise statements about convergence, it is usually necessary to consider how the time to convergence scales with the system size. For instance, one would like to know how fast, as a function of the lattice size, a given dynamical process on a lattice converges to its steady state. 

\begin{figure}[t]
\centering
\subfigure[~cutoff]{\includegraphics[scale=0.45]{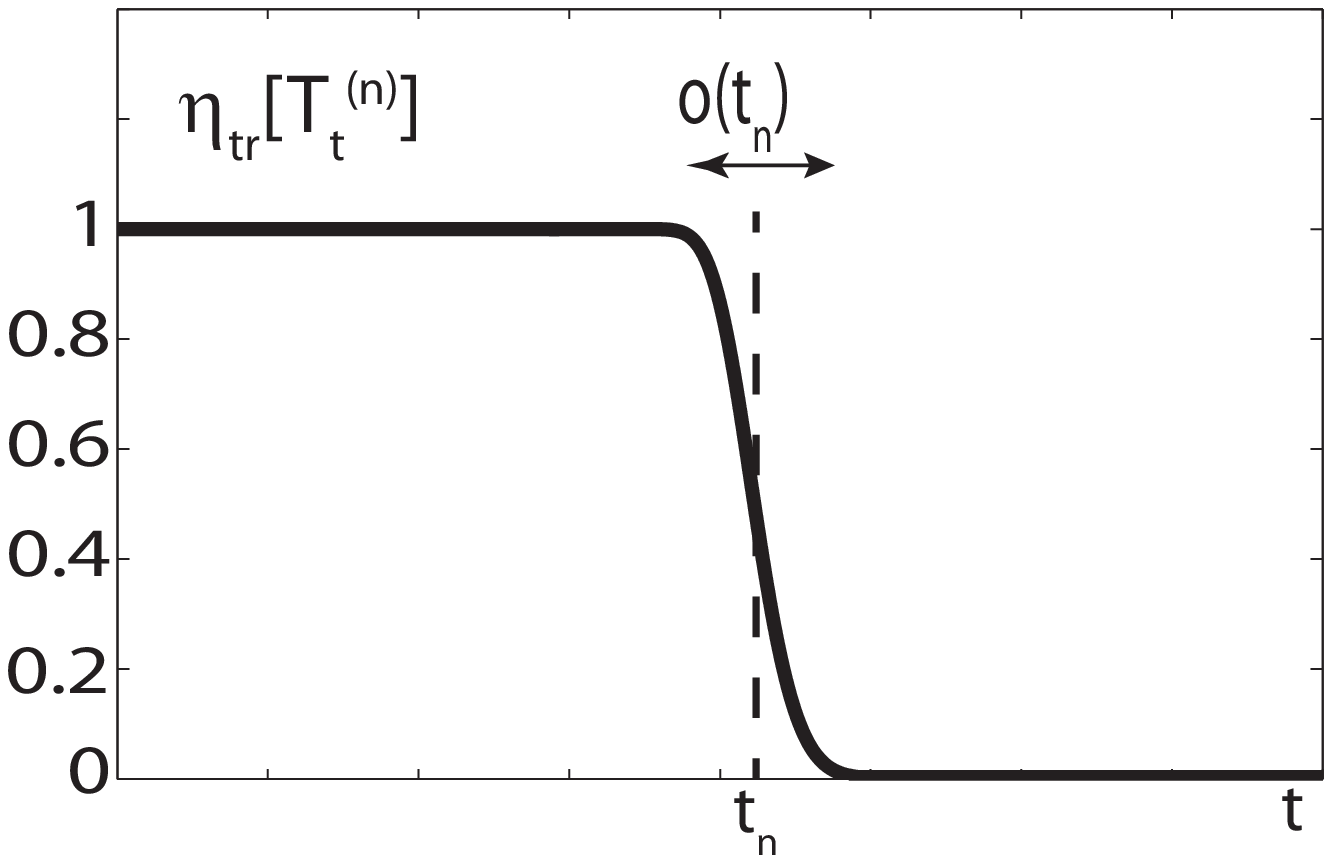}\label{fig:subfig1}}
\subfigure[~pre-cutoff]{\includegraphics[scale=0.45]{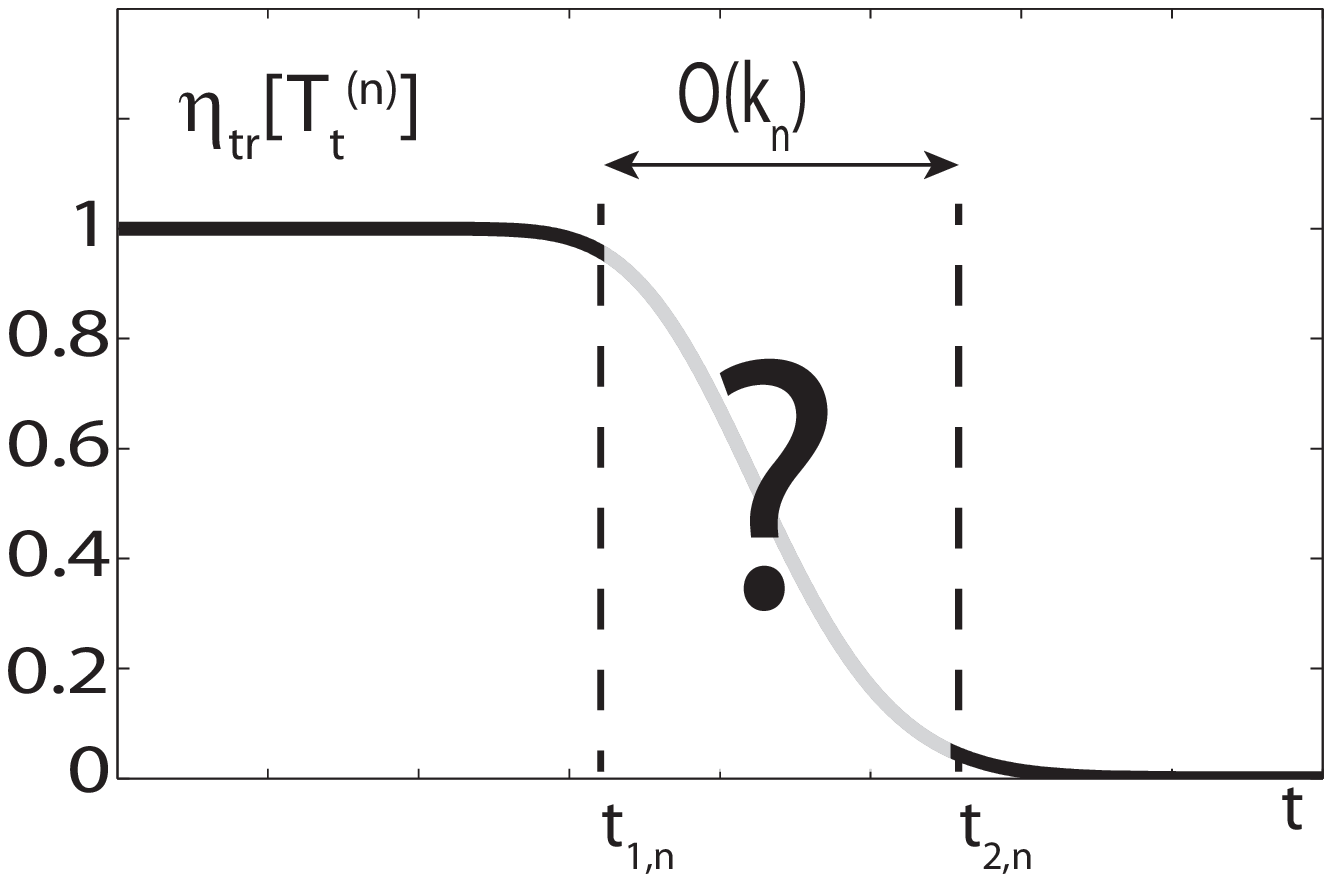}\label{fig:subfig2}}
\label{Cutoff Phenomenon}
\caption{\label{figurecutoff}\subref{fig:subfig1} Behavior of the contraction $\eta_\tr[T_t^{(n)}]$ as a function of evolution time $t$ for a one-parameter semigroup of channels that exhibits a cutoff (for large system size $n$). At time $t=t_n$, essentially all dependence on the initial state vanishes in a small window of time $o(t_n)$. \subref{fig:subfig2} In the case of pre-cutoff, the contraction will be close to $1$ for times $t<t_{1,n}=\Theta(k_n)$ and close to $0$ for $t>t_{2,n}=\Theta(k_n)$, but now the gap between $t_{1,n}$ and $t_{2,n}$ may be of order $O(k_n)$. }
\end{figure}

It was observed a while ago in the setting of classical Markov chains, that for a special set of chains this question can be answered exactly when the size of the system becomes large. This behavior, which has been coined the \textit{Cutoff Phenomenon} \cite{cutoff1,cutoff2,cutoff3}, characterizes the situation when for some (possibly long) period of time some information from the initial state is perfectly preserved until a critical time. Shortly after this critical time, however, essentially no information of the initial state can be recovered from the time-evolved state anymore. For large system sizes $n$, the contraction of the channel as a function of time $t$ will look like a step function at the \textit{cutoff time} $t_n$ (see Fig.~\ref{figurecutoff}a).

This behavior has been observed and proved to occur in a number of interesting examples of classical Markov chains. One case where this phenomenon is particularly pronounced, and which triggered widespread popular interest, is in card shuffling, where it was shown that a deck of $n$ cards is well mixed after exactly $3/2 \log{n}$ riffle shuffles, and poorly mixed under $3/2 \log{n}$ riffle shuffles (when $n$ becomes large)  \cite{cutoff2}. In particular, this guarantees casino owners that if their dealers riffle shuffle their 52 poker cards seven or more times before each draw, then they do not need to worry about about players trying to improve their odds by counting cards\footnote{The punchline in this example is that the cutoff behavior can be exploited to perform a spectacular magic trick called ``Premo'' \cite{CardTrick}, where the magician can guess a card chosen randomly by a person from the audience who afterwards shuffles the deck a few times. The information in the initial configuration (i.e., the identity of the card that the person put on top) is preserved and can be identified before a ``time'' of exactly $\frac{3}{2} \log{n}$ riffle shuffles (for large $n$).}. Several other processes have also been shown to exhibit cutoffs, including random walks on graphs with a discrete group structure, birth and death type chains, and some Monte Carlo sampling methods \cite{cutoff3,birthdeathcut,MCMCcut}. We now give a formal definition in the quantum setting:
\begin{definition}[Cutoff]\label{Cutoff}Let $T^{(n)}_t$ be a sequence, indexed by the ``system size'' $n$, of one-parameter semigroups of quantum channels. We say that $T^{(n)}_t$ exhibits a cutoff (in trace-norm) at times $t_n$, if for any real $c>0$:
\bea c<1\quad&\Rightarrow&\quad\lim_{n\rightarrow \infty} \eta_\tr[T^{(n)}_{c t_n}] = 1~,\nonumber\\
c>1\quad&\Rightarrow&\quad\lim_{n\rightarrow \infty} \eta_\tr[T^{(n)}_{c t_n}] = 0~.\nonumber\eea
\end{definition}
We point out that this is not the only definition of a cutoff, and in a sense it is an incomplete one, as it does not provide detailed information about the cutoff window, i.e.~the width of the drop-off at $t_n$ (see Fig.~\ref{figurecutoff}a).

In many situations, it is difficult to prove an actual cutoff, whereas it might be easier to show a weaker statement which gives only the precise order of magnitude of the time to convergence:
\begin{definition}[Pre-cutoff]With the same assumptions as in Definition \ref{Cutoff}, we say that $T^{(n)}_t$ exhibits a pre-cutoff (in trace-norm) of order $\Theta(k_n)$ for some sequence $k_n$, if there exist times $t_{1,n}$ and $t_{2,n}$ that are both of order $\Theta(k_n)$, such that 
\bea c<1\quad&\Rightarrow&\quad\lim_{n\rightarrow \infty} \eta_\tr[T^{(n)}_{c t_{1,n}}] = 1~,\nonumber\\
c>1\quad&\Rightarrow&\quad\lim_{n\rightarrow \infty} \eta_\tr[T^{(n)}_{c t_{2,n}}] = 0~.\nonumber\eea
\end{definition}

We again emphasize that the cutoff phenomenon does not cover all types of pre-asymptotic behavior. For example, a process can follow a polynomial decay for a certain amount of time before it settles into the asymptotic (exponential) regime. We also point out that the Cutoff Phenomenon can be defined with respect to any monotone distance measure (even unbounded ones like the $\chi^2$-divergence). A cutoff in one distance measure does not imply a cutoff in another measure. This suggests that obtaining such a tight contraction estimate might in some cases reveal information about only one specific facet of the convergence behavior, associated with the given distance measure. For instance, the trace-norm contraction measure of a channel tells us whether a single bit of classical information is preserved after a certain time, but makes no statement about the entire amount of information or the preservation of quantum information.

\section{Main Results}\label{mainresultssecdt}

In this section, we present two situations which exhibit behavior related to the Cutoff Phenomenon introduced above: commuting Liouvillians and, as a more restrictive but still relevant class, Liouvillians acting independently on subsystems. In the latter situation, the contraction behavior of the constituent channels is known, and we ask how the contraction of the tensor product of channels behaves. We show in general terms that a sequence of tensor product channels exhibits a pre-cutoff of order $\Theta(\log{n})$, where $n$ is the number of tensor factors. In the next section, we discuss some specific situations of this kind where an actual cutoff occurs, which includes the dissipative preparation of stabilizer states.

Our first theorem provides a general upper bound on the contraction of a channel from a one-parameter semigroup whose Liouvillian is composed of commuting parts, i.e.~$\cL=\sum_j \cL_j$ where $[\cL_j,\cL_k]=0$. In this case, the gap of the full Liouvillian is at least the minimum of the gaps of its constituent parts. We show that in this context the time to convergence is upper bounded by $O(\log{n})$ times the convergence time of the ``slowest'' constituent channel, where $n$ is the number of commuting terms in the Liouvillian.

\begin{theorem}[Contraction for commuting Liouvillians]\label{CommLiouv}Let $\cL_j:\cM_d\rightarrow\cM_d$ be Liouvillians which commute, i.e.~$[\cL_j,\cL_k]=0$ for all $j,k=1,\ldots,n$. Define $\cL\equiv\sum_j \cL_j$, and the corresponding semigroups of channels $T_{t,j}\equiv e^{t\cL_j}$ and $T_t\equiv e^{t\cL}$ $(t\geq0)$. Then:
\be\eta_\tr[T_t]~\leq~\sum_j \eta_\tr[T_{t,j}]~.\ee
\end{theorem}
\proof{
The theorem is proved by induction. Let $T_{\varphi,1}$ be the projector onto the asymptotic space of $T_{t,1}$, and let $T_{\varphi,j\neq1}$ be the projector onto the asymptotic space of $T_{t,j\neq1}\equiv e^{t\sum_{j\neq1}\cL_j}$, see Eqn.~(\ref{periphchannel}). Note that $T_t=T_{t,1}T_{t,j\neq1}$ and $T_\varphi=T_{\varphi,1} T_{\varphi,j\neq1}$, by commutativity of the Liouvillians. Then,
\bea\eta_\tr[T_t] &=& \half\sup_{\rho \in \cS_d} ||(T_{t,1} T_{t,j\neq1}-T_{\varphi,1} T_{\varphi,j\neq1})(\rho)||_1\nonumber\\
&=& \half\sup_{\rho \in \cS_d} ||T_{t,1} (T_{t,j\neq1}-T_{\varphi,j\neq1})(\rho)+(T_{t,1}-T_{\varphi,1})( T_{\varphi,j\neq1})(\rho)||_1\nonumber\\
&\leq& \eta_\tr[T_{t,1}] + \eta_\tr[T_{t,{j\neq1}}]~,\label{lastaddivitveupperbound}\eea
where the last inequality follows from the triangle inequality, from monotonicity of the trace-norm under quantum channels, and by definition of $\eta_\tr[T_t]$. By induction, we get $ \eta_\tr[T_t]\leq \sum_j \eta_\tr[T_{t,j}]$. \qed}

An immediate consequence of Theorem \ref{CommLiouv}, also using Theorem \ref{contractionrate}, is that for a system described by $n$ commuting Liouvillians with bounded gaps, the convergence time will be upper bounded by $O(\log{n})$. Note however, that commuting Liouvillians should not be confused with classical processes. Indeed, as described in Section \ref{graphstatessect}, graph states can be prepared dissipatively as stationary states of commuting Liouvillians, whereas these states can be highly entangled (e.g.~cluster state).


\bigskip
 
Our second result gives the general contraction behavior of a tensor power of a one-parameter semigroup of quantum channels. This is a special case of commuting Liouvillians, but where strict upper and lower bounds can be derived as the number of tensor factors becomes large, thereby establishing a pre-cutoff as defined in Section \ref{cutoffsection}.
\begin{theorem}[Pre-cutoff for tensor powers]\label{cutoffThm}Let $\cL:\cM_d\rightarrow\cM_d$ be a Liouvillian with gap $\bar{\lambda}$, and let $T_t\equiv e^{t\cL}$ $(t\geq0)$. The sequence of one-parameter semigroups $T^{(n)}_t\equiv T^{\otimes n}_t$ exhibits a pre-cutoff in trace-norm at times $t_{1,n}=\log{(n)}/2\bar{\lambda}$ and $t_{2,n}=\log{(n)}/\bar{\lambda}$.
\end{theorem}
\proof{
Here and below we use the fact that $(T\otimes S)_\varphi=T_\varphi\otimes S_\varphi$ for any pair of quantum channels $T$ and $S$. To prove the lower bound, let $c\in(0,1)$:
\bea \eta_\tr[T^{(n)}_{ct_{1,n}}] &=& \sup_{\rho\in\cS_{d^{n}}} d_\tr\big(T^{(n)}_{ct_{1,n}}(\rho),T^{(n)}_{\varphi}(\rho)\big)~\geq~\sup_{\sigma\in\cS_{d}} d_\tr\big((T_{ct_{1,n}}(\sigma))^{\otimes n},(T_{\varphi}(\sigma))^{\otimes n}\big) \label{LB1}\nonumber\\
&\geq& 1-\exp{\left[-\half n \sup_{\sigma\in\cS_{d}} d_\tr^2\left(T_{ct_{1,n}}(\sigma),T_{\varphi}(\sigma)\right)\right]} \label{LB2}\nonumber\\
&\geq& 1-\exp{\left[-\frac{L^2}{2} n e^{-2ct_{1,n}\bar{\lambda}}\right]}~=~1-\exp{\left[-\frac{L^2}{2} n^{1-c}\right]}\,\,\rightarrow\,1~~~(n\to\infty)~.\label{LB3}\eea
The first inequality is obtained by restricting the supremum to product states $\rho=\sigma^{\otimes n}$, the next from Lemma \ref{TensorIneq} (see below), and the last follows from Theorem \ref{contractionrate} (with some constant $L>0$). Hence, $\lim_{n \rightarrow \infty} \eta_\tr[T^{(n)}_{ct_{1,n}}] = 1$, for $c\in(0,1)$.

For the upper bound, we apply Theorem \ref{CommLiouv} to get $\eta_\tr[T^{(n)}_{t}]\leq n\eta_\tr[T_{t}\otimes \id_{n-1}]$, where ${\rm id}_{n-1}$ is the identity channel on $n-1$ sites. In the following paragraph we show $\eta_\tr[T_{t}\otimes \id_{n-1}]\leq4d\eta_\tr[T_{t}]$. Now, for any given $c>1$ one can choose $\nu<{\bar{\lambda}}$ such that $c\nu/{\bar{\lambda}}>1$, and by Theorem \ref{contractionrate} one can find $R$ such that $\eta_\tr[T_{t}]\leq Re^{-\nu t}$ for all $t\geq0$. Combining all this, we finally get that for any $c>1$,
\be\eta_\tr[T^{(n)}_{ct_{2,n}}]~\leq~4dn\,Re^{-\nu ct_{2,n}}~=~4Rdn^{1-c\nu/{\bar{\lambda}}}\,\rightarrow\,0~~~(n\to\infty)~.\ee

It remains to show that $\eta_\tr[T\otimes\id]\leq4d\eta_\tr[T]$ for any channel $T:\cM_d\rightarrow\cM_d$ and any identity channel $\id:\cM_{d'}\rightarrow\cM_{d'}$. The inequality (used below) between the norm $||\cdot||_{1-1}$ on superoperators induced by the trace-norm and its stabilized version $||\cdot||_{cb}$ is proven in \cite{paulsen} (exercise 3.11). In the following, $X\in\cM_d\otimes\cM_{d'}$ and $A+iB\in\cM_d$ denote arbitrary matrices, $A$ and $B$ are Hermitian, and $P,Q\in\cM_d$ are positive semidefinite with $PQ=0$. Further note that $||A||_1=||(A+iB)+(A-iB)||_1/2\leq(||A+iB||_1+||A-iB||_1)/2=||A+iB||_1$, and similarly $||B||_1\leq||A+iB||_1$. Thus:
\bea\eta_\tr[T\otimes\id]&=&\sup_{\rho\in\cS_{dd'}}d_\tr\left(T\otimes\id(\rho),T_\varphi\otimes\id(\rho)\right)~\leq\sup_{||X||_1\leq1}\frac{1}{2}||((T-T_\varphi)\otimes\id)(X)||_1\nonumber\\
&\leq&\frac{1}{2}||T-T_\varphi||_{cb}~\leq~\frac{d}{2}||T-T_\varphi||_{1-1}~=~\frac{d}{2}\sup_{||A+iB||_1\leq1}||(T-T_\varphi)(A+iB)||_1\nonumber\\
&\leq&d\sup_{||A||_1\leq1}||(T-T_\varphi)(A)||_1~=~d\sup_{||P-Q||_1\leq1}||(T-T_\varphi)(P-Q)||_1\nonumber\\
&\leq&2d\sup_{||P||_1\leq1}||(T-T_\varphi)(P)||_1~=~4d~\eta_\tr[T]~.\nonumber\eea\qed}

Theorem \ref{cutoffThm} can be generalized to certain cases where $T^{(n)}_t$ is the tensor product of a set of one-parameter semigroups $T^i_t$ that are not all identical. See \cite{BLY} for the analogous classical result.

Theorem \ref{cutoffThm} establishes pre-cutoff rather than actual cutoff behavior. But at the end of subsection \ref{amplitudedampingsection} we show examples where, for any chosen $r\in[1,2]$, a cutoff occurs at times $t_n=\log(n)/r{\bar{\lambda}}$. This means that $t_{1,n}$ and $t_{2,n}$ in Theorem \ref{cutoffThm}, viewed as upper and lower bounds on the contraction, are tight when expressed in terms of the gap ${\bar{\lambda}}$.

\bigskip

The following Lemma completes the proof of Theorem \ref{cutoffThm} (cf.~Eqn.~(\ref{LB3})); the inequalities (\ref{BuresTensorProd}) for the Bures distance will be used to show cutoff in Proposition \ref{convergenceCorr}. For collections $\rho_i,\,\sigma_i \in \cS_d$ of density matrices ($i=1,...,n$), define $\rho^{(n)}\equiv \bigotimes_{i=1}^n\rho_i$ and similarly $\sigma^{(n)}$.
\begin{lemma}[Distances between tensor product states]\label{TensorIneq}Let $\rho_i,\sigma_i \in \cS_d$, $i=1,...,n$, and denote by $d_\tr$, $d_B$ the trace and Bures distances, respectively. Then the following inequalities hold:
\be1-\exp{\left[-\sum_{i=1}^n d^2_B(\rho_i,\sigma_i)\right]}~\leq~d^2_B(\rho^{(n)},\sigma^{(n)})~\leq~\sum_{i=1}^n d^2_B(\rho_i,\sigma_i)~.\label{BuresTensorProd}\ee
\be1-\exp{\left[-\frac{1}{2}\sum_{i=1}^n d^2_\tr(\rho_i,\sigma_i)\right]}~\leq~d_\tr(\rho^{(n)},\sigma^{(n)})~\leq~\sum_{i=1}^n d_\tr(\rho_i,\sigma_i)~.\label{TNTensorProd}\ee
\end{lemma}
\proof{
The fidelity is multiplicative under tensor products, $F(\rho^{(n)},\sigma^{(n)})=\prod_{i=1}^nF(\rho_i,\sigma_i)$. Also, by induction it is easily seen that $(1-\prod_i x_i)\leq \sum_i (1-x_i)$ for any collection of reals  $x_i\in[0,1]$. Thus:
\be d^2_B(\rho^{(n)},\sigma^{(n)})~=~1-\prod_{i=1}^nF(\rho_i,\sigma_i)~\leq~\sum_{i=1}^n (1-F(\rho_i,\sigma_i))~=~\sum_{i=1}^n d^2_B(\rho_i,\sigma_i)~.\nonumber\ee

Since $\prod_i e^{x_i-1} \geq \prod_i x_i$ whenever $x_i\geq0$, we get the lower bound in (\ref{BuresTensorProd}):
\be d^2_B(\rho^{(n)},\sigma^{(n)})~=~1-\prod_{i=1}^nF(\rho_i,\sigma_i)~\geq~1- \prod_{i=1}^n e^{F(\rho_i,\sigma_i)-1}~=~1-\exp{\left[-\sum_{i=1}^n d^2_B(\rho_i,\sigma_i)\right]}~.\nonumber\ee

The upper bound in (\ref{TNTensorProd}) follows by a calculation similar to the one yielding Eqn.~(\ref{lastaddivitveupperbound}), while the lower bound follows from (\ref{BuresTensorProd}) and two of the inequalities in (\ref{TrNrmFid}).\qed}

\section{Examples of the cutoff phenomenon and applications}\label{examplessect}

Theorem \ref{cutoffThm} establishes a pre-cutoff and thereby estimates, up to a factor of 2, the time to convergence. The next natural question is: when does an actual cutoff occur? We discuss two such situations. The first concerns tensor powers of primitive channels where the input states are restricted to be separable, the second concerns tensor powers of channels whose unique fixed point is a pure state. In subsection C we give an explicit example of this situation (the qubit amplitude damping channel), and in subsection D we apply this to the dissipative preparation of graph states and discuss how the cutoff phenomenon determines the exact convergence time of this process.

\subsection{Primitive Channels with Separable Initial States}

Beyond Theorem \ref{cutoffThm}, we can establish a sharp cutoff for primitive Liouvillians when the inputs are restricted to be fully separable quantum states between the $n$ channels:
\be\cS_{\bigotimes_id_i}^{\rm sep}~:=~\Big\{\,\sum_kp_k\rho_1^k\otimes\ldots\otimes\rho_n^k\,\big|\,p_k\geq0,\,\sum_kp_k=1,\,\rho_i^k\in\cS_{d_i}\,\Big\}~.\ee
\begin{proposition}[Primitive Liouvillians with separable inputs]\label{convergenceCorr}
Let $\cL:\cM_d\rightarrow\cM_d$ be the generator, with gap ${\bar{\lambda}}$, of a one-parameter semigroup of primitive channels $T_t\equiv e^{t\cL}$ $(t\geq0)$, and define the trace-norm contraction of $T_t^{(n)}\equiv T_t^{\otimes n}$ restricted to separable input states:
\be\eta_\tr^{\rm sep}[T^{(n)}_t]~:=~\sup_{\rho \in \cS^{\rm sep}_{d^{\otimes n}}} d_\tr\left(T^{\otimes n}_t(\rho),T_\varphi^{\otimes n}(\rho)\right)~.\label{defsepinpcontr}\ee
Then, the sequence of one-parameter semigroups $T^{(n)}_t\equiv T^{\otimes n}_t$ exhibits a cutoff (with respect to the contraction measure $\eta_\tr^{\rm sep}$) at times $t_n=\log{(n)}/2\bar{\lambda}$.
\end{proposition}
\proof{
From the primitivity of the channel we get that $T_\varphi(\rho)=\sigma$ for any input state $\rho\in\cS_d$, where $\sigma$ is the unique stationary state of $\cL$. Further, as $\sigma$ is of full rank, the new bound from Proposition \ref{burestnLinear} and bounds from Eqn.~(\ref{TrNrmFid}) together with Theorem \ref{contractionrate} show that, for any $\nu<\bar{\lambda}$, there exist constants $R>L>0$ such that $Le^{-t\bar{\lambda}}\leq \eta_B[T_t]\leq Re^{-t\nu}$ for all $t\geq0$.

The rest of the proof follows the same lines as the proof of Theorem \ref{cutoffThm}. However, we first show the theorem here for the Bures metric, i.e.~replacing $d_\tr$ in (\ref{defsepinpcontr}) by $d_B$. For proving the lower bound, the same arguments as the ones leading to Eqn.~(\ref{LB3}) show that $\lim_{n \rightarrow \infty}\eta^{\rm sep}_B[T_{ct_n}] = 1$ for $c\in(0,1)$. For the upper bound, note that due to convexity of the Bures distance (derived from concavity of the fidelity \cite{NielsenChuang}) the supremum in (\ref{defsepinpcontr}) is reached for a product state $\rho=\bigotimes_{i=1}^n\rho_i$, so that Lemma \ref{convergenceCorr} can be applied:
\bea\left(\eta_B^{\rm sep}[T^{(n)}_{ct_n}]\right)^2&=&\sup_{\rho_i\in \cS_{d}} d_B^2\Big(\bigotimes_{i=1}^nT_{ct_n}(\rho_i),\,\sigma^{\otimes n}\Big)~\leq~\sum_{i=1}^n\sup_{\rho_i\in\cS_d}d_B^2(T_{ct_n}(\rho_i),\sigma)\nonumber\\
&\leq&nR^2e^{-2c\nu t_n}~=~R^2n^{1-c\nu/{\bar{\lambda}}}~.\nonumber\eea
As in the proof of Theorem \ref{cutoffThm}, for each given $c>1$ one can choose $\nu$ and $R$ accordingly to show $\lim_{n \rightarrow \infty}\eta^{\rm sep}_B[T_{ct_n}]=0$, which proves a cutoff at times $t_n$. Finally, by Eqn.~(\ref{TrNrmFid}), a cutoff in the Bures contraction $\eta_B^{\rm sep}$ is equivalent to a cutoff in the trace-norm contraction $\eta_\tr^{\rm sep}$, at the same times $t_n$.\qed}
We do not know whether the separable input assumption is actually necessary for Proposition \ref{convergenceCorr}. If the assumption were indeed necessary, then the statement  would imply the possibility of increased storage time of classical information due to an entangled encoding.

\subsection{Channels with Unique Pure State Fixed Point}

\begin{proposition}[Unique pure state fixed point]\label{convergencePure}Suppose that the pure state $\psi=\ket{\psi}\bra{\psi}\in\cS_d$ is the unique stationary state of the Liouvillian $\cL$, which has gap ${\bar{\lambda}}$ and generates the channels $T_t:=e^{t\cL}$ $(t\geq0)$. Then, $T^{(n)}_t\equiv T^{\otimes n}_t$ exhibits a trace-norm cutoff at times $t_n=\log{(n)}/\bar{\nu}$, for some $\bar{\lambda}\leq\bar{\nu}\leq2\bar{\lambda}$.
\end{proposition}
\proof{
Since $\cL$ has only one stationary state, the peripheral spectrum of $e^{t\cL}$ is trivial for all $t>0$, so that $T_\varphi(\rho)=\psi$ for all $\rho\in\cS_d$. This, together with well-known inequalities relating the fidelity and the trace-norm between a pure and a mixed state \cite{NielsenChuang}, yields:
\bea1-\inf_{\rho\in\cS_{d^n}} F^2(T_t^{\otimes n}(\rho),\psi^{\otimes n})~\leq~\eta_{tr}[T_t^{\otimes n}]~\leq~\sqrt{1-\inf_{\rho\in\cS_{d^n}} F^2(T_t^{\otimes n}(\rho),\psi^{\otimes n})}\label{upperlowerbound}~.\eea
The last infimum can be evaluated explicitly in the case at hand:
\bea\label{firstF}\inf_{\rho\in\cS_{d^n}} F^2(T_t^{\otimes n}(\rho),\psi^{\otimes n})&=&\inf_{\rho\in\cS_{d^n}}\tr{[T_t^{\otimes n}(\rho)\psi^{\otimes n}]}~=~\inf_{\rho\in\cS_{d^n}}\tr{[\rho\,\left(T_t^*(\psi)\right)^{\otimes n}]}\label{lastinf}\nonumber\\
&=&\lambda_{min}\left(\left(T_t^*(\psi)\right)^{\otimes n}\right)~=~\left[\,\lambda_{min}\left(T_t^*(\psi)\right)\,\right]^n\nonumber\\
&=&\left[\,1-\lambda_{max}\left(\1-T_t^*(\psi)\right)\,\right]^n~=~\big(\,1-||T_t^*(\1-\psi)||_\infty\,\big)^n\label{finalinftynorm}~,\nonumber\eea
where in the last step we used that $T_t$ is trace-preserving ($T_t^*(\1)=\1$) and that $T_t^*$ is a positive map. Thus, from equation (\ref{upperlowerbound}) above:
\bea1-\left(1-||T_t^*(\1-\psi)||_\infty\right)^n~\leq~\eta_{tr}[T_t^{\otimes n}]~\leq~\sqrt{1-\left(1-||T_t^*(\1-\psi)||_\infty\right)^n}\label{finalresult}~.\eea

Since $T_\varphi(\rho)=\psi$ for all $\rho\in\cS_d$, we have $T_\varphi(A)=\psi\tr[A]$ for all $A\in\cM_d$, so that its dual is given by $T^*_\varphi(B)=\1\tr[B\psi]$ for $B\in\cM_d$. Thus:
\bea||T_t^*(\1-\psi)||_{\infty}~=~||\,(T_t^*-T_{\varphi}^*)(\psi)\,||_\infty~.\nonumber\eea
Now we consider this last equation in a similar way as in the proof of Theorem \ref{contractionrate} in Appendix \ref{appendixb}: When one writes $\psi$ as a linear combination of the generalized eigenvectors of $\cL^*$, then considering large times $t$ will essentially pick out the Jordan-eigenvalue(s) occurring in $\psi$ which has largest real part $-\bar{\nu}<0$ (i.e., not the eigenvalue 0), and among these it will pick out the polynomial(s) of highest degree $J\geq0$ that are occupied (i.e., occur with non-zero coefficient in the linear decomposition of $\psi$). This means that, for any arbitrarily chosen $t_0>0$, there exist constants $0<C_1\leq C_2$ such that
\bea C_1(\bar{\nu} t)^J e^{-\bar{\nu} t}~\leq~||(T_t^*-T_\varphi^*)(\psi)||_\infty~\leq~C_2(\bar{\nu} t)^J e^{-\bar{\nu} t}\qquad\forall t>t_0~.\nonumber\eea

Using this in Eqn.~(\ref{finalresult}) gives
\bea1-\left(1-C_1(\bar{\nu} t)^J e^{-\bar{\nu} t}\right)^n~\leq~\eta_{tr}[T_t^{\otimes n}]~\leq~\sqrt{1-\left(1-C_2(\bar{\nu} t)^J e^{-\bar{\nu} t}\right)^n}\label{finally}~.\nonumber\eea
This proves a cutoff for $T_t^{\otimes n}$ at times $t_n=(\log n)/{\bar{\nu}}$, since for any constants $c,K>0$ and $J\geq0$:
\bea\lim_{n\ra\infty}1-(1-K(\bar{\nu}c t_n)^Je^{-\bar{\nu} ct_n})^n&=&1-\lim_{n\ra\infty}\left(1+\frac{-K(c\log{n})^Jn^{1-c}}{n}\right)^n\nonumber\\
&=& 1-\lim_{n\ra\infty}\exp{\left(-K(c\log{n})^Jn^{1-c}\right)}~=~\left\{\begin{array}{ll}1\,,&c<1\\0\,,&c>1~.\end{array}\right.\nonumber\eea

As $(-{\bar{\nu}})$ is the real part of an eigenvalue of $\cL$, it is evident that ${\bar{\nu}}\geq{\bar{\lambda}}$, and Theorem \ref{cutoffThm} shows ${\bar{\nu}}\leq2{\bar{\lambda}}$.\qed}

Both infima in the upper bound and in the lower bound in Eqn.~(\ref{upperlowerbound}) are attained for some pure \emph{product} state $\rho=\varphi^{\otimes n}$, even though this is not clear for the supremum that achieves $\eta_{tr}[T_t^{\otimes n}]$. The paradigmatic example in the following subsection saturates the upper bound $\bar{\nu}=2\bar{\lambda}$, but we also provide modifications of this example where $\bar{\nu}$ takes on any values between $\bar{\lambda}$ and $2\bar{\lambda}$.

\subsection{Qubit Amplitude Damping}\label{amplitudedampingsection}

The amplitude damping process (on qubits) describes the situation where the excited state $\ket{1}$ decays into the ground state $\ket{\psi}:=\ket{0}$ at a constant rate $\gamma$. This corresponds to a Master equation with a single Lindblad operator $L:=\sqrt{\gamma}\ket{0}\bra{1}$ and no coherent contribution:
\be\cL(\rho)~:=~L\rho L^\dagger-\frac{1}{2}L^\dagger L\rho-\frac{1}{2}\rho L^\dagger L~=~\gamma\left(\ket{0}\bra{0}\cdot\bra{1}\rho\ket{1}-\frac{1}{2}\ket{1}\bra{1}\rho-\frac{1}{2}\rho\ket{1}\bra{1}\right)~.\label{amplitudedecayL}\ee

A straightforward calculation shows that:
\be\label{etaex}\eta_{tr}[e^{t\cL}]~=~\left\{\begin{array}{lll}
e^{-\gamma t}&,\,0\leq t\leq(\log2)/{\gamma}~~&(\text{attained for}~\rho=\ket{1})\\
e^{-\gamma t/2}/\sqrt{4(1-e^{-\gamma t})}&,\,t\geq(\log2)/{\gamma}~~&(\text{attained for}~\rho\propto{\ket{1}+\sqrt{1-2e^{-\gamma t}}\ket{0}})\,.\end{array}\right.\nonumber\ee
Thus, $\eta_{tr}[e^{t\cL}]$ decays asymptotically in time as $e^{-\gamma t/2}$, and \emph{not} as $e^{-\gamma t}$, which one would expect from the analogous classical noise process (the classical Markov map has eigenvalues $0$ and $-\gamma$, whereas the Liouvillian (\ref{amplitudedecayL}) has two additional eigenvalues $-\gamma/2$).

By Proposition \ref{convergencePure}, the semigroups $(e^{t\cL})^{\otimes n}$ exhibit a trace-norm cutoff at times $t_n=(\log n)/{\bar{\nu}}$ for some ${\bar{\nu}}$ with $\bar{\lambda}=\gamma/2\leq\bar{\nu}\leq2\bar{\lambda}=\gamma$. ${\bar{\nu}}$ can be computed explicitly by using $||T_t^*(\1-\psi)||_\infty=e^{-\gamma t}$ in Eqn.~(\ref{finalresult}), or from the proof of Proposition \ref{convergencePure} by writing $\psi$ as a linear combination of eigenvectors of $\cL^*$:
\be\psi~=~\ket{0}\bra{0}~=~\1-\ket{1}\bra{1}~,\label{lstardecomp}\ee
where $\1$ and $\ket{1}\bra{1}$ are eigenvectors of $\cL^*$ with eigenvalues $0$ and $-\gamma$, respectively. Thus, ${\bar{\nu}}=\gamma=2{\bar{\lambda}}$, and the cutoff occurs at times $t_n=(\log n)/\gamma$.

For system size $n$ it thus takes time $O(\log n)$ before convergence happens, even though the Liouvillians
\be\cL^{(n)}~=~\cL\otimes\id\otimes\ldots\otimes\id\,+\,\id\otimes\cL\otimes\ldots\otimes\id\,+~\ldots~+\,\id\otimes\id\otimes\ldots\otimes\cL~,\label{locallindbladians}\ee
which generate the semigroups $T_t^{(n)}=e^{t\cL^{(n)}}$, have a gap ${\bar{\lambda}}^{(n)}={\bar{\lambda}}=\gamma/2$ which is \emph{independent} of $n$. Therefore, this example refutes the conventional wisdom whereby ``the gap governs the convergence time''.

If, in addition to the Liouvillian (\ref{amplitudedecayL}), there are also processes with Lindblad operators $\sqrt{\alpha}\ket{0}\bra{0}$ and $\sqrt{\beta}\ket{1}\bra{1}$ acting on each qubit, then the steady state $\psi$ and its decomposition (\ref{lstardecomp}) into eigenvectors of the dual evolution operator are as above (in particular, ${\bar{\nu}}=\gamma$), and a cutoff occurs at times $t_n=(\log n)/\gamma$. In this new situation, however, the gap is given by ${\bar{\lambda}}=\min\{\gamma,(\gamma+\alpha+\beta)/2\}$, which shows that the bounds on the cutoff time given by Proposition \ref{convergencePure} and implied by Theorem \ref{cutoffThm} are tight.

\subsection{Dissipative Preparation of Graph States}\label{graphstatessect}

As an application of the results from subsections B and C, we consider the dissipative preparation of graph states. This task was considered in \cite{DSP1,DQC1}, where it was shown that a set of local Lindblad operators can be constructed in a way analogous to Eqn.~(\ref{locallindbladians}), such that the unique stationary state of the process is the desired graph state and that the spectral gap of the process is independent of the number $n$ of particles or stabilizer operators.

We complete this analysis by showing that the convergence time, as measured by the trace-norm contraction, scales as $\log{n}$ with the system size. The trace-norm contraction is the relevant quantity to consider in this case, as it quantifies the maximal failure probability when the graph state is used for further quantum information processing, like the cluster state for measurement-based quantum computing. We actually show that the dissipative preparation of graph states exhibits a cutoff (in trace-norm) at times of order $O(\log{n})$ in the sense of Definition \ref{Cutoff}. Although still efficient, the $\log{n}$ scaling of the preparation time again indicates that the gap does not fully determine the convergence behavior.
\begin{proposition}[Dissipative preparation of graph states]\label{graphstate}
Any graph state on $n$ sites, associated to a graph of maximal degree $k$, can be prepared dissipatively in a time of order $\log n$ by using $n$ Lindblad operators that are at most $k$-local.

In fact, for large $n$, the preparation procedure described in \cite{DSP1,DQC1} takes exactly time $t_n=(\log n)/\gamma$ to converge to the desired graph state, where $\gamma$ is the decay rate ($\gamma/2=\,$spectral gap) of each local Lindblad operator and when starting from the most disadvantageously chosen initial state.
\end{proposition}
\proof{
Given a set $\{S_k\}_{k=1}^n$ of stabilizer operators, the unique state which is an eigenstate of $S_k$ with eigenvalue $+1$ for every $k$  is called a stabilizer state. Graph states \cite{stabilizer} are a special case of these and can be described by an undirected graph with $n$ vertices. The stabilizer operators of the graph state are then $S_k=\sigma^x_k \prod_{j\in {\rm nbhd} (k)} \sigma^z_j$, where nbhd$(k)$ denotes the set of all vertices connected to vertex $k$ by an edge.

The stabilizer operators of a graph state uniquely define a ``graph basis'', written as $\{\ket{\Phi_{i_1,...,i_n}}\}_{i_l\in\{0,1\}}$, by $S_k\ket{\Phi_{i_1,...,i_n}}=(-1)^{i_k}\ket{\Phi_{i_1...,i_n}}$. These basis vectors satisfy $\sigma^z_k\ket{\Phi_{i_1,...,i_k=1,...,i_n}}=\ket{\Phi_{i_1,...,i_k=0,...,i_n}}$, and the ``graph state'' is $\ket{\Phi_{0,...,0}}$.

Define the $n$ Lindblad operators \cite{DSP1} ($k=1,\ldots,n$)
\be L_k~=~\sqrt{\gamma}\,\sigma^z_k\frac{\1-S_k}{2}~,\label{GSLind}\ee
and observe that $L_k \ket{\Phi_{i_1,...,i_k=1,...,i_n}}=\sqrt{\gamma}\ket{\Phi_{i_1,...,i_k=0,...,i_n}}$ and $L_k\ket{\Phi_{i_1,...,i_k=0,...,i_n}}=0$. Thus, in the graph basis, each of these Lindblad operators acts as one term of the sum (\ref{locallindbladians}) acts in the computational basis. Therefore, together they act like the tensor product of amplitude damping channels in subsection C, now with the graph state as the stationary state. Proposition \ref{convergencePure} or, more explicitly, subsection C thus prove a cutoff at times $(\log n)/\gamma$ for the preparation of graph states.\qed}

Note in particular, Proposition \ref{graphstate} shows that, for the procedure described by Eqn.~(\ref{GSLind}), there exist some initial states for which one can guarantee convergence not to occur before time $(\log n)/\gamma$.

\section{Conclusion}\label{conclusionsect}

In this paper, we have introduced the notion of the Cutoff Phenomenon in the context of quantum information theory and applied it to analyze the convergence behavior of some composite quantum processes in continuous time. In particular, we show that the convergence, measured in the trace-norm, of a tensor product of one-parameter semigroups of time evolutions always exhibits cutoff-type behavior. We identify two specific cases (primitive channels with separable initial states, and channels with a unique pure fixed point), which exhibit a true cutoff. We provided a full analysis of the problem of dissipatively preparing graph states, and show that the convergence time scales as $\log{n}$ in the number of stabilizer elements $n$. 

Finally, we conclude by noting two directions where the methods introduced in this paper could be of use. The first is the task of passive error protection in the presence of local noise. It was shown recently \cite{QMem1} that, if the noise is locally depolarizing and allowing for arbitrary Hamiltonian control, an optimal protection time of order $O(\log{n})$ can be achieved. Theorem \ref{cutoffThm} gives a strict upper bound on the amount of time that one bit of classical (and hence also quantum) information can be encoded into $n$ qubits, when every qubit is subjected to local noise, and no Hamiltonian control is allowed for. The upper bound happens to coincide with the one in \cite{QMem1}, indicating that their result might not be restricted to depolarizing channels, but could be a general feature of tensor product channels. Along similar lines, a second extension of the above results is in the study of continuous time quantum information theory, where channels are replaced by one-parameter semigroups, and standard objects, such as channel capacities and compression rates, become functions of time.

\bigskip

{\it Acknowledgments}: MJK thanks Frank Verstraete for bringing the beautiful cutoff results of Diaconis to his attention. We acknowledge financial support from the European projects COQUIT and QUEVADIS, the CHIST-ERA/BMBF project CQC, and the Niels Bohr International Academy, and from the Alfried Krupp von Bohlen und Halbach-Stiftung.


\appendix

\section{Proof of Proposition \ref{burestnLinear}}\label{appendixa}
We now give a proof of Proposition \ref{burestnLinear} from Section \ref{distancemeasuresection}.

\proof{
Denoting by $\lambda_i$ and $\ket{i}$ the eigenvalues and -vectors of $\sigma$ ($i=1,\ldots,d$), it is shown in \cite{huebner} that
\be d_B^2(\rho,\sigma)~=~\frac{1}{4}\sum_{i,j=1}^d\frac{\left|\,\bra{i}(\rho-\sigma)\ket{j}\,\right|^2}{\lambda_i+\lambda_j}\,+\,O\left(d_\tr^3(\rho,\sigma)\right)~,\nonumber\ee
using that all norms on a finite-dimensional vector space are equivalent. The last term is upper bounded by $Kd_\tr^3(\rho,\sigma)$ with some constant $K=K(\sigma)>0$, and we bound the denominator by $2\lambda_{min}(\sigma)$:
\bea d_B^2(\rho,\sigma)&\leq&\frac{1}{4}\sum_{i,j=1}^d\frac{\bra{i}(\rho-\sigma)\ket{j}\bra{j}(\rho-\sigma)\ket{i}}{2\lambda_{min}(\sigma)}\,+\,Kd_\tr^3(\rho,\sigma)~=~\frac{||\rho-\sigma||_2^2}{8\lambda_{min}(\sigma)}+Kd_\tr^3(\rho,\sigma)\nonumber\\
&\leq&\frac{||\rho-\sigma||_1^2}{16\lambda_{min}(\sigma)}+Kd_\tr^3(\rho,\sigma)~=~\left(\frac{1}{4\lambda_{min}(\sigma)}+Kd_\tr(\rho,\sigma)\right)d_\tr^2(\rho,\sigma)~,\nonumber\eea
where we used $||\rho-\sigma||_2\leq||\rho-\sigma||_1/\sqrt{2}$ for the traceless matrix $\rho-\sigma$. To upper bound the last term in parentheses by $1/\lambda_{min}(\sigma)$, we set $\epsilon:=3/(4K\lambda_{min}(\sigma))$ and the claim follows.\qed}

{\bf Remark:} A general \emph{linear} upper bound of the form $d_B(\rho,\sigma)\leq Cd_\tr(\rho,\sigma)$ as in Eqn.~(\ref{burestrlineareqn}) cannot hold for $\sigma\notin\cS^+_d$. For instance, in this case there exist density matrices $\rho\in\cS_d$ orthogonal to $\sigma$ ($\rho\sigma=0$). Then defining $\rho_\delta:=\delta\rho+(1-\delta)\sigma$ (for $\delta\in[0,1]$) one has $d_\tr (\rho_\delta,\sigma)=\delta$ and $F(\rho_\delta,\sigma)=\sqrt{1-\delta}$, so that
\be d_\tr(\rho_\delta,\sigma)~\leq~2d^2_B(\rho_\delta,\sigma)\qquad\forall\delta\in[0,1]~.\nonumber\ee
Thus, there do not exist constants $C,\epsilon>0$ such that $d_B(\rho,\sigma)\leq Cd_\tr(\rho,\sigma)$ holds for all $d_\tr(\rho,\sigma)<\epsilon$.

\section{Proof of Theorem \ref{contractionrate}}\label{appendixb}
This appendix proves Theorem \ref{contractionrate} from Section \ref{distancemeasuresection}.

\proof{
Let $\hat{\cL}$ be the matrix representation of the Liouvillian. The Jordan normal form gives
\be\hat{\cL}~=~\hat{S}\,\,\bigoplus_j\hat{J}_j(\lambda_j)\,\,\hat{S}^{-1}\nonumber\ee
for some invertible matrix $\hat{S}$, where $\lambda_j$ are the eigenvalues of $\cL$, and $\hat{J}_j(\lambda_j)$ are Jordan blocks of the following form (note that eigenvalues $\lambda_j$ with ${\rm Re}(\lambda_j)=0$ have one-dimensional Jordan blocks, so in particular all $0$ eigenvalues):
\be\hat{J}_j(\lambda_j)~=~\left( \begin{array}{cccc} \lambda_j & \lambda_j & 0  &  \\   & \lambda_j & \lambda_j & 0 \\   &  & &\ldots \end{array} \right)~.\nonumber\ee
Let $d_j\geq1$ be the dimension of Jordan block $j$. Then, in the Jordan basis $\{\ket{k}\}$ defined by this,
\be e^{t\hat{J}_j(\lambda_j)}~=~e^{t\lambda_j}\,\sum_{l=1}^{d_j}\sum_{k=1}^l\,\frac{(t\lambda_j)^{l-k}}{(l-k)!}\,\ket{k}\bra{l}~.\label{exponentiation}\ee
Let $||\cdot||$ be the operator norm, and denote by $\kappa(\hat{S}):=||\hat{S}||\,||\hat{S}^{-1}||$ the condition number of the similarity transformation into Jordan form. Then:
\bea\big|\big|e^{t\hat{\cL}}-\hat{T}_\varphi\big|\big|&=&\Big|\Big|\hat{S}\,\bigoplus_{j:{\rm Re}\lambda_j\neq0}  e^{t\hat{J}_j(\lambda_j)}\,\hat{S}^{-1}\Big|\Big|~\leq~\kappa(\hat{S})\,e^{-t\bar{\lambda}}\,\max_{j:{\rm Re}(\lambda_j)\neq0}\,\left|\left|\sum_{l=1}^{d_j}\sum_{k=1}^l \frac{(t\lambda_j)^{l-k}}{(l-k)!}\ket{k}\bra{l}\right|\right|\nonumber\\
&\leq&\kappa(\hat{S})\,e^{-t\bar{\lambda}}\max_{j: {\rm Re}(\lambda_j)\neq0}\left(\sum_{l=1}^{d_j}\sum_{k=1}^l \frac{(t|\lambda_j|)^{l-k}}{(l-k)!}\right)~\leq~Ce^{-t\bar{\lambda}}\max\big\{(t\bar{\lambda})^{J-1},1\big\}~,\nonumber\eea
where $C$ is a $t$-independent constant and $J:=\max_j d_j$ is the dimension of the largest Jordan block. The last step is obtained by factoring $(t\bar{\lambda})^{J-1}$ out of the sum (for times $t\geq1/{\bar{\lambda}}$) and bounding the remaining term by a constant. Thus clearly, for any $\nu<\bar{\lambda}$ there exists a constant $K>0$ such that the last expression is upper bounded by $Ke^{-t{\bar{\lambda}}}$ (for all $t\geq0$).

The lower bound is obtained similarly (letting $\hat{J}_1(\lambda_1)$ be any Jordan block with ${\rm Re}(\lambda_1)=-{\bar{\lambda}}$):
\bea\big|\big|e^{t\hat{\cL}}-\hat{T}_\varphi\big|\big|&\geq&\kappa(\hat{S})^{-1}\,\Big|\Big|\bigoplus_{j:{\rm Re}(\lambda_j)\neq0}e^{t\hat{J}_j(\lambda_j)}\Big|\Big|~=~\kappa(\hat{S})^{-1}\,\max_{j:{\rm Re}(\lambda_j)\neq0}\,\Big|\Big|e^{\hat{J}_j(\lambda_j)}\Big|\Big|\nonumber\\
&\geq&\kappa(\hat{S})^{-1}\,\Big|\Big|e^{t\hat{J}_1(\lambda_1)}\Big|\Big|~\geq~\kappa(\hat{S})^{-1}\,\max_{1\leq k\leq l\leq d_1}\Big|\left(e^{t\hat{J}_1(\lambda_1)}\right)_{kl}\Big|~\geq~\kappa(\hat{S})^{-1}\,e^{-t\bar{\lambda}}~,\nonumber\eea
where maximum in the second-to-last expression runs over all matrix elements $(k,l)$ in (\ref{exponentiation}), and in the last step we chose $k=l=1$.

Using these upper and lower bounds on $||\hat{T}_t-\hat{T}_\varphi||$, we invoke Lemma \ref{linopcontr} (below) to complete the proof of Theorem \ref{contractionrate}, by lumping all of the time-independent constants together and denoting them by $L>0$ and $R$.\qed}

\begin{lemma}\label{linopcontr}
Let $T: \cM_d \rightarrow \cM_d$ be a quantum channel, and $||\cdot||$ the operator norm. Then:
\be\frac{1}{8\sqrt{d}}\,||\hat{T}-\hat{T}_\varphi||~\leq~\eta_\tr[T]~\leq~\frac{\sqrt{d}}{2}\,||\hat{T}-\hat{T}_\varphi||~.\ee
\end{lemma}
\proof{
For the lower bound, we use in the first inequality below that every $X\in\cM_d$ can be written as $X=P_1-P_2+iP_3-iP_4$ with positive semidefinite $P_i$ satisfying $P_1P_2=P_3P_4=0$, and that then $||P_j||_2^2\leq\sum_i||P_i||_2^2=||X||_2^2$. In the following chains we also use that $||X||_2\leq||X||_1\leq\sqrt{d}||X||_2$.
\bea||\hat{T}-\hat{T}_\varphi||&=&\sup_{||X||_2\leq1}||(T-T_\varphi)(X)||_2~\leq~4\sup_{P\geq0,||P||_2\leq1}||(T-T_\varphi)(P)||_2\nonumber\\
&\leq&4\sup_{P\geq0,||P||_1\leq\sqrt{d}}||(T-T_\varphi)(P)||_1~=~4\sqrt{d}\sup_{P\geq0,\tr[P]\leq1}||(T-T_\varphi)(P)||_1~=~8\sqrt{d}\eta_\tr[T]~.\nonumber\eea
\bea\eta_\tr[T]&\leq&\frac{1}{2}\sup_{||X||_1\leq1}||(T-T_\varphi)(X)||_1~\leq~\frac{1}{2}\sup_{||X||_2\leq1}\sqrt{d}||(T-T_\varphi)(X)||_2~=~\frac{\sqrt{d}}{2}||\hat{T}-\hat{T}_\varphi||~.\nonumber\eea\qed}

\end{document}